\documentclass[draft]{agujournal2019}
\usepackage{url} 
\usepackage{lineno}
\usepackage[inline]{trackchanges} 
\usepackage{soul}
\usepackage{amsmath,amssymb,bm,slashbox,units}
\usepackage{textcomp}

\draftfalse

\journalname{Earth and Space Science}

\makeatletter
\providecommand*{\shuffle}{%
  \mathbin{\mathpalette\shuffle@{}}%
  }
\newcommand{\define}{\stackrel{\text{def}}{=}}
  \newcommand*{\shuffle@}[2]{%
  \sbox0{$#1\vcenter{}$}%
  \kern .15\ht0 
  \rlap{\vrule height .25\ht0 depth 0pt width 2.5\ht0}%
  \raise.1\ht0\hbox to 2.5\ht0{%
    \vrule height 1.75\ht0 depth -.1\ht0 width .17\ht0 %
    \hfill
    \vrule height 1.75\ht0 depth -.1\ht0 width .17\ht0 %
    \hfill
    \vrule height 1.75\ht0 depth -.1\ht0 width .17\ht0 %
    }%
    \kern .15\ht0 
    }
\makeatother
\begin{document}

%
%


\title{
Machine learning technique 
using the signature method for automated quality control 
of the Argo profiles}
%

%
%




\authors{Nozomi Sugiura\affil{1} and Shigeki Hosoda\affil{1}}


\affiliation{1}{Research and Development Center for Global Change, JAMSTEC, Yokosuka, Japan}



\correspondingauthor{Nozomi Sugiura}{nsugiura@jamstec.go.jp}




\begin{keypoints}
\item Machine learning for the quality control flags of  Argo profiles 
was performed.
\item By converting each profile sequence of temperature, salinity, and pressure 
into its signature, classification was performed efficiently.
\item The signature is regarded as a fundamental object that represents a data sequence.
\end{keypoints}

%
%

%
%


\begin{abstract}
A profile from the Argo ocean observation array
is a sequence of three-dimensional vectors
composed of pressure, salinity, and temperature,
appearing as a continuous curve in three-dimensional space. 
The shape of this curve is faithfully represented by 
a path signature, which is a collection of all the iterated integrals. 
Moreover, the product of two terms of the signature of a path
can be expressed as the sum of higher-order terms.
Thanks to this algebraic property,
a nonlinear function of profile shape
can always be represented by a weighted linear combination
of the iterated integrals,
which enables machine learning of 
a complicated function of the profile shape. 
In this study, we performed supervised learning for 
existing Argo data with quality control flags 
by using the signature method, 
and demonstrated the estimation performance by cross-validation.
Unlike rule-based approaches,
which require several complicated and possibly subjective rules,
this method is simple and objective in nature 
because it relies only on past knowledge regarding the 
shape of profiles.
This technique should be critical to realizing 
automatic quality control for Argo profile data. 
\end{abstract}


%
%


\section{Introduction} \label{Intro}
Argo is an international effort collecting
high-quality temperature and salinity profiles,
typically from the upper 2000 m of the global
ocean \cite{doi:10.1029/2004EO190002}. 
The data come from battery-powered autonomous
floats that drift mostly at a depth where
they are stabilized at a constant pressure level.
At typically 10-day intervals, the floats rise to the surface
for approximately 6 h while measuring temperature
and salinity. On surfacing, the satellites position
the floats and receive the transmitted data.
Now, the array of over 3000 floats provides 
100,000 temperature/salinity profiles annually
distributed over the global oceans at an average 3-degree spacing. 
The quality control (QC) of the massive 
Argo profile data \cite{argo2019} must be systematic to 
keep the quality of the observational data homogeneous
and to utilize human resources efficiently.
In addition, accurately quantifying the relationship between
the profile shape and the effect it has on oceanic processes 
is essential for understanding the ocean state 
through the profile observation.
Conventionally, significant time and effort are spent 
assigning the quality control flag to each Argo profile.

Regarding attempts for advanced automatic QC procedures on oceanographic profiles, 
    some studies have been applied to the Argo CTD profile because of the huge amount of data accumulated for 200 
    million profiles over 20 years. For example, \shortciteA{Udaya2013} provided a semi-automatic QC procedure 
    using objective mapping to remove anomalous values from the profiles. \shortciteA{UDAYABHASKAR2017469} demonstrated
    automatic QC by defining the convex fulls from
    the climatological dataset. 
    Meanwhile, \citeA{Ono2015} attempted to apply a machine learning method to  
    the delayed-mode QC of Argo profiles 
    towards a possible automatic QC system for an Argo data stream. 
    Similarly, an integrated Argo data flow using machine 
    learning was introduced to be an automated system with an improved QC ability 
    (presented by \shortciteA{Maze2017} in the report of 
    the 18th Argo Data Management Meeting). 
    Thus,
    QC procedures for oceanographic data have been 
    gradually improved by many researchers using advanced tools or methods.

The discrimination procedures involved in the automation of  the quality control 
have been performed mainly in a rule-based manner \cite<e.g.>[]{7344896,7838290,7849862}.
As an alternative and more flexible approach,
this study attempted to automate the process 
via supervised learning of the human judgment process.
In doing so, it is essential to quantify the profile shape 
so that the function that yields the quality control flag 
can be expressed as a linear combination of the numerical values that represent the profile shape.
The machine learning thereby reduces to 
a linear optimization problem that can be easily solved.
The key tool that enables this quantification is 
the signature, which is the set of all iterated integrals
\cite{chevyrev2016primer,levin2013learning},
proposed in the theory of rough path by \citeA{lyons2007differential}.

In this research, we propose a procedure of 
first converting the vector sequence of each Argo profile into 
a sequence of real numbers that represents its shape and then 
expressing a nonlinear function of the shape 
in the form of a linear combination of these numbers;
this conversion facilitates machine learning of the nonlinear function.
A machine learning experiment regarding the function was performed
and applied to automatic assignment of quality control flags 
to the profiles.

\section{Theoretical background}\label{TB}
The central concept in this study is the signature,
proposed in the theory of rough path by \citeA{lyons2007differential}.
In what follows, we briefly introduce the concept of signature and 
the notation used in this paper.
For more details, refer to \citeA{chevyrev2016primer},

As perceived from a re-examination of 
controlled differential equations (refer \ref{Pic}),
characteristics of a data sequence can be represented 
by the signature, which comprises the iterated integrals.
Note, in this paper, the subscript notation $X_{\tau}$ 
is used to denote dependence on the parameter $\tau \in [0,t]$;
$A^{\bullet n}$ and $A^{\otimes n}$ denote the $n$-th power and
$n$-times tensor product, respectively,
but otherwise, a superscript denotes a component.

Suppose we have a sequence of $d$-dimensional vectors $X_u~(0 \leq u \leq t)$.
Let the time order be $0<t_1< \cdots < t_n<t$.
We define the iterated integral for indices 
$i_1, \cdots i_n = 1, \cdots, d$ as 
\begin{linenomath*}\postdisplaypenalty=0
	   \begin{align}
	    \mathrm{X}^{(i_1 \cdots i_n)} &= \int_{t_n=0}^t \cdots \int_{t_1=0}^{t_2} dX^{i_1}_{t_1}
	     \cdots dX^{i_n}_{t_n},
	   \end{align}
\end{linenomath*}
where we should be careful about the difference 
between the font 
for a sequence of vectors $X^k_{t_1}$
and the one for an iterated integral $\mathrm{X}^{(k_1k_2\cdots k_n)}$.
By treating all the index values together, we obtain
a tensor of order $n$:
\begin{linenomath*}\postdisplaypenalty=0
  \begin{align}
   \mathbf{X}_n&=\int_{0<t_1< \cdots < t_n<t}
   dX_{t_1}\otimes    \cdots \otimes   dX_{t_n}\quad n=1,2,\cdots,\label{IteratedI}
  \end{align}
\end{linenomath*}
and $\mathbf{X}_0$ is constant $1$.
Moreover, by putting together the iterated integrals 
for all combinations of the indices, we obtain the signature up to degree $n$:
\begin{linenomath*}\postdisplaypenalty=0
\begin{align}
		\mathcal{S}^n(X) &=\left(\mathbf{X}_0, \mathbf{X}_1, \mathbf{X}_2, \cdots, \mathbf{X}_n\right),
\end{align}
\end{linenomath*}
which has $(d^{n+1}-1)/(d-1)$ components.
For instance, the signature up to degree $2$ for a $2$-dimensional sequence is
\begin{linenomath*}\postdisplaypenalty=0
\begin{align}
\mathcal{S}^2(X) 
&=
\left(\mathrm{X}^{()},
\begin{bmatrix}
\mathrm{X}^{(1)}\\
\mathrm{X}^{(2)}
\end{bmatrix},
\begin{bmatrix}
\mathrm{X}^{(11)}&\mathrm{X}^{(12)}\\
\mathrm{X}^{(21)}&\mathrm{X}^{(22)}
\end{bmatrix}
\right)
=
\left(1, 
\begin{bmatrix}
X_{0,t}^1\\
X_{0,t}^2
\end{bmatrix},
\begin{bmatrix}
\frac12 (X_{0,t}^1)^{\bullet 2}& 
\int_0^t X_{0,u}^1 dX_{u}^2\\
\int_0^t X_{0,u}^2 dX_{u}^1
&\frac12 (X_{0,t}^2)^{\bullet 2}
\end{bmatrix}
\right),
\end{align}
\end{linenomath*}
where $X_{0,t}^i \define X_t^i-X_0^i$, $\bullet 2$ denotes the second power,
and $\mathrm{X}^{()}=\mathbf{X}_0=1$.
Note that the order of integrands matters in
 $\int_0^t X_{0,u}^1 dX_{u}^2$ and $\int_0^t X_{0,u}^2 dX_{0,u}^1$.
In general, it is important for the signature to encode the order in which  
each component changes along the path.

Suppose we have two paths,
 $X=\{X_{\tau}\}_{0\leq\tau\leq s}$ and  $Y=\{Y_{\tau}\}_{s\leq\tau\leq t}$.
Their concatenation is the path defined by
\begin{linenomath*}\postdisplaypenalty=0
\begin{align}
\left( X*Y\right)_{\tau}=
\begin{cases}
X_{\tau} & \text{if~} \tau \in [0,s]\\
X_s+Y_{\tau}-Y_s & \text{if~} \tau \in [s,t].
\end{cases}
\label{concate}
\end{align}
\end{linenomath*}
On the other hand, regarding their signatures,
$\mathcal{S}^n(X)=(\mathbf{X}_0, \mathbf{X}_1, \cdots, \mathbf{X}_n)$ and
	       $\mathcal{S}^n(Y)=(\mathbf{Y}_0, \mathbf{Y}_1, \cdots, \mathbf{Y}_n)$,
we can define the product as
\begin{linenomath*}\postdisplaypenalty=0
\begin{align}
\mathcal{S}^n(X)\otimes\mathcal{S}^n(Y)&\define(\mathbf{Z}_0, \mathbf{Z}_1, \cdots, \mathbf{Z}_n),\label{prod_sig} \\
\mathbf{Z}_h &=\sum_{k=0}^h \mathbf{X}_k\otimes \mathbf{Y}_{h-k},
\end{align}
\end{linenomath*}
whose components are
	   \begin{linenomath*}\postdisplaypenalty=0
	    \begin{align}
	    \mathrm{Z}^{(i_1\cdots i_h)}
	    &=
	    \sum_{k=0}^h
	    \mathrm{X}^{(i_1 \cdots i_k)}
	    \mathrm{Y}^{(i_{k+1} \cdots i_h)}. \label{prod}
	    \end{align}
	   \end{linenomath*}
For instance, the product 
      of the signatures, up to degree $2$, for the $2$-dimensional sequence is
	   \begin{linenomath*}\postdisplaypenalty=0
	    \begin{align}
\mathcal{S}^2(X)\otimes \mathcal{S}^2(Y)
        &=\left( \mathrm{Y}^{()},
        \begin{bmatrix}
          \mathrm{Y}^{(1)}+\mathrm{X}^{(1)} \\
          \mathrm{Y}^{(2)}+\mathrm{X}^{(2)} 
        \end{bmatrix},
        \begin{bmatrix}
          \mathrm{Y}^{(11)}+\mathrm{X}^{(1)}\mathrm{Y}^{(1)}+\mathrm{X}^{(11)}&
          \mathrm{Y}^{(12)}+\mathrm{X}^{(1)}\mathrm{Y}^{(2)}+\mathrm{X}^{(12)}\\
          \mathrm{Y}^{(21)}+\mathrm{X}^{(2)}\mathrm{Y}^{(1)}+\mathrm{X}^{(21)}&
          \mathrm{Y}^{(22)}+\mathrm{X}^{(2)}\mathrm{Y}^{(2)}+\mathrm{X}^{(22)}
        \end{bmatrix} \right). \label{prod2}
	    \end{align}
	   \end{linenomath*}

In this manner, the set of signatures 
has a group structure in the free tensor algebra
with respect to the product $\otimes$.
Furthermore, 
Chen's identity \cite{Chen1958}:
\begin{linenomath*}\postdisplaypenalty=0
\begin{align}
\mathcal{S}^n(X*Y)
=
\mathcal{S}^n(X)\otimes \mathcal{S}^n(Y)\label{chen1}
\end{align}
\end{linenomath*}
is satisfied, which
defines
a homomorphism from path space with concatenation (\ref{concate}) to
signature space with group operation (\ref{prod_sig}).

In the context of geophysics, we can show that 
some diagnoses for oceanographic conditions
are written in terms of iterated integrals. 
Consider a vertical sequence of vector $(P_{\tau},S_{\tau},T_{\tau})$ 
(pressure, salinity, and temperature) 
in the ocean.
     \begin{enumerate}
      \item The first-order iterated integrals are
\begin{linenomath*}\postdisplaypenalty=0
	    \begin{align*}
	     \mathrm{X}^{(P)} &= \int_{\tau=0}^t dP_{\tau} =P_{t}-P_{0},\quad
	     \mathrm{X}^{(S)} = S_{t}-S_{0}, \quad
	     \mathrm{X}^{(T)} = T_{t}-T_{0},
	    \end{align*}
\end{linenomath*}
	    which are profile depth, sea surface salinity, and sea surface temperature, respectively.
      \item 
	    The second-order iterated integrals include
\begin{linenomath*}\postdisplaypenalty=0
	    \begin{align*}
	     \mathrm{X}^{(PP)} &= \frac12 (P_{t}-P_{0})^{\bullet 2},\quad
	     \mathrm{X}^{(SP)} = \int_{\tau=0}^t (S_{\tau}-S_{0}) dP_{\tau},\quad
	     \mathrm{X}^{(TP)} = \int_{\tau=0}^t (T_{\tau}-T_{0}) dP_{\tau},
	    \end{align*}
\end{linenomath*}
	    which represent the square of profile depth,
	    total salinity content, and total heat content, respectively.
\end{enumerate}
We find another example in \ref{da}.

Note that $P$ is treated equally to $S,T$ in the above 
    because the seemingly redundant parameter $\tau$ is essential to ensure that 
    the path has no self intersection 
    and the signature is invariant under the reparameterization of $\tau$.
    If one parameterizes $T,S$ with $P$, the path would be drawn 
    on a two-dimensional $T,S$-surface,
    which loses considerable information on the shape of the sequence.

\section{Method}\label{sec_method}
The data used
in this research were observed by the global array of Argo floats \cite{argo2019},  
each of which floats and sinks from the sea surface to a depth of approximately $2000 \mathrm{m}$.

\begin{figure}
  \begin{center}
   \includegraphics[width=42em]{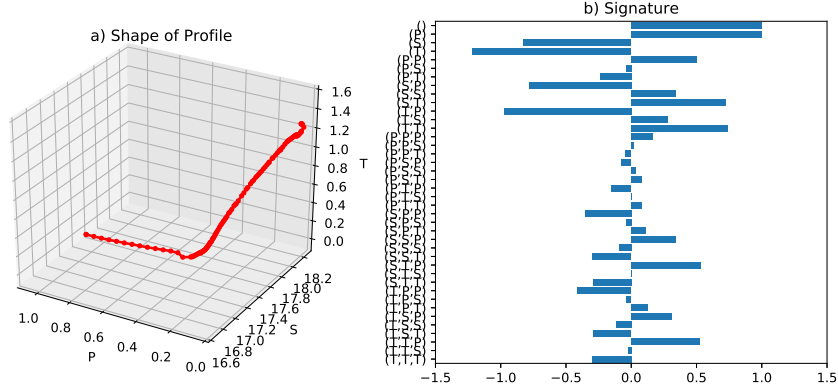}
\caption {An example of a) profile shape obtained from Argo observation,
and b) their first few iterated integrals.
For example, $(T,P)$ denotes the iterated integral 
$\mathrm{X}^{(TP)} = \int_{0}^t \int_{0}^{t_2} dT_{t_1} dP_{t_2}$.
$P, S$, and $T$ are divided in advance
by $2000\unit{dbar}$, $2\unit{psu}$, and $20$\textdegree C,  respectively, 
and thus dimensionless.
    \label{argo_profile1}}
  \end{center}
\end{figure}

\begin{figure}
  \begin{center}
   \includegraphics[width=42em]{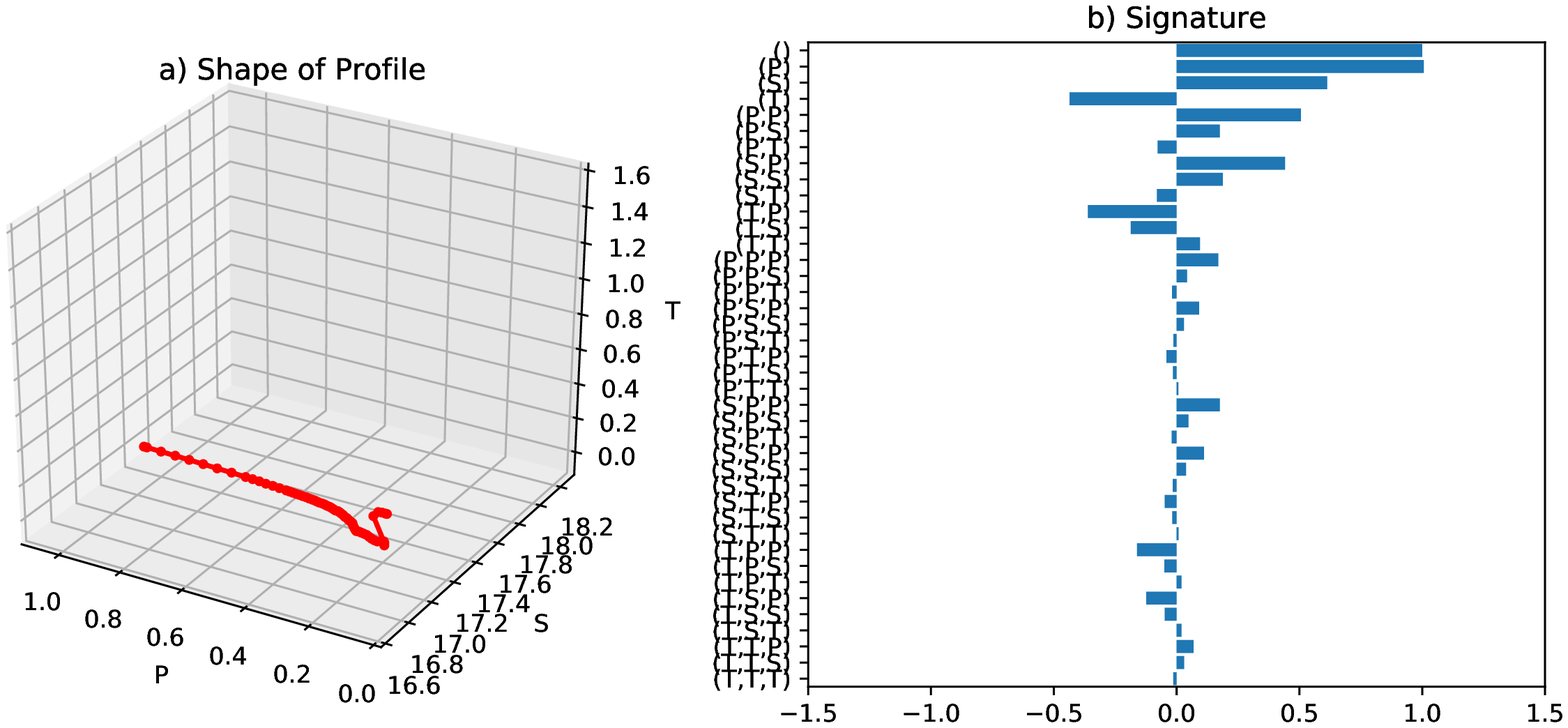}
\caption {Another example of a) profile shape obtained from Argo observation,
and b) their first few iterated integrals.
    \label{argo_profile2}}
  \end{center}
\end{figure}

Because the shape of a vector sequence $(P_{\tau},S_{\tau},T_{\tau})$ 
    is only perceived in a certain reference frame,
    it is convenient to make  the original quantities dimensionless;
    $P$ in $\unit{dbar}$, $S$ in $\unit{psu}$, and $T$ in \textdegree{}C
    into $\widehat{P}=P/2000$, $\widehat{S}=S/2$, and 
    $\widehat{T}=T/20$,
    where divisor $(2000,2,20)$ is chosen as a typical scale of the components.
    For simplicity, henceforth, we omit the hat symbol for the component.
Figures \ref{argo_profile1} and \ref{argo_profile2} show 
examples of the vertical profiles
of temperature, salinity, and pressure,
along with the corresponding 
iterated integrals.
By virtue of quality control procedures with manual judgment,
the quality control flags are already assigned to all of the data.

Here, we describe the basic concept of the signature method
 and how to apply it to Argo profiles.
We also 
explain how to construct a procedure for 
supervised learning using the signature and how to verify the results.

  \subsection{Representing the Argo profile shape by signature}
   \subsubsection{Computation of signature}
Suppose we have $d$-dimensional profile data $\{X_{\tau}\}_{0 \le \tau \le t}$
that can be seen as
a line graph connecting points $X_0=X_{u_1},X_{u_2},\cdots,X_{u_L}=X_t$;
then, we can compute its iterated integrals as follows:
     \begin{enumerate}
      \item 
	   For line segment 
$X_{\text{seg}}\define \{X^i_{u}+X^i_{u, u'}\tau\}
_{0 \le \tau \le 1}^{i=1,2,\cdots,d}$, which has starting point 
	    $X^i_{u}$ and  slope $X^i_{u, u'}$, 
the iterated integrals are calculated as 
	    \begin{linenomath*}\postdisplaypenalty=0
	     \begin{align}
	     \mathrm{X}^{(i)} &= X^i_{u, u'}, \qquad
	     \mathrm{X}^{(ij)} = \frac{1}{2!} X^i_{u, u'}X^j_{u, u'} , \nonumber\\
	     \mathrm{X}^{(ijk)} &= \frac{1}{3!} X^i_{u, u'} X^j_{u, u'}X^k_{u, u'}, \label{unit}
	     \end{align}
	    \end{linenomath*}
	    and the $0$-th iterated integral is constant $1$.
	    In this case, the signature (up to degree $n$) is nothing but a 
	    commutative exponential function for the vector $X_{u, u'}$:
	    \begin{linenomath*}\postdisplaypenalty=0
	     \begin{align}
	    \mathcal{S}^n(X_{\text{seg}})&=
	    \sum_{h=0}^{n}
	    \frac{1}{h!}
	    \sum_{i_1,  i_2,  \cdots,  i_h}
	    \prod_{k=1}^h X^{i_k}_{u, u'}\mathbf{e}_{i_k},
	     \end{align}
	    \end{linenomath*}
	    where $\mathbf{e}_{i_k}$ is the $i_k$-th unit vector.
      \item
Let the time order be $ s \leq u \leq t$.
By concatenating a path $X_{s, u}$ 
from time $s$ to $u$ with
a path $X_{u, t}$ from time $u$ to $t$, 
we obtain
a path $X_{s, t}$ from time $s$ to $t$,
whose signature is the product of 
the signatures:
	    \begin{linenomath*}\postdisplaypenalty=0
	     \begin{align}
	    \mathcal{S}^n(X_{s, t})  &=  \mathcal{S}^n(X_{s, u})  \otimes \mathcal{S}^n(X_{u, t}),
	     \label{chen}
	     \end{align}
	    \end{linenomath*}
which is due to Chen's identity (\ref{chen1}).
      \item
By concatenating the paths successively using Eq.\,(\ref{chen}),
we can compute
the signature for the whole line graph.
     \end{enumerate}
     The numerical computation of the signature
     in this study is performed by using Python library {\tt Esig} \cite{esig}.
   \subsubsection{Lead-lag transformation\label{sec_ll}}
Suppose we have a sequence of $d_0$-dimensional  ($d_0=3$) vectors
with length $L+1$:
\begin{linenomath*}\postdisplaypenalty=0
       \begin{align*}
	X &= \left( X_0,  X_1,  \cdots,  X_L \right) =
	\left(
	\begin{bmatrix}
	 P_0\\	 S_0\\	 T_0
	\end{bmatrix}, 
	\begin{bmatrix}
	 P_1\\	 S_1\\	 T_1
	\end{bmatrix}, 
       \cdots, 
	\begin{bmatrix}
	 P_L\\	 S_L\\	 T_L
	\end{bmatrix}
	\right)
       \end{align*}
\end{linenomath*}
To more precisely grasp the shape of the line graph, we perform 
a lead-lag transformation \cite{chevyrev2016primer},
which defines a
sequence of $d(=2d_0)$-dimensional  vectors with length $L \cdot d +1$:
\begin{linenomath*}\postdisplaypenalty=0
\begin{align*}
	\begin{bmatrix}
	 \color{blue}{P_0}\\	 \color{blue}{S_0}\\	 \color{blue}{T_0}\\	 \color{blue}{P_0}\\	 \color{blue}{S_0}\\	 \color{blue}{T_0}
	\end{bmatrix}, &
	\begin{bmatrix}
	 \color{red}{P_1}\\	 S_0\\	 T_0\\	 P_0\\	 S_0\\	 T_0
	\end{bmatrix}, 
	\begin{bmatrix}
	 P_1\\	 \color{red}{S_1}\\	 T_0\\	 P_0\\	 S_0\\	 T_0
	\end{bmatrix}, 
	\begin{bmatrix}
	 P_1\\	 S_1\\	 \color{red}{T_1}\\	 P_0\\	 S_0\\	 T_0
	\end{bmatrix}, 
	\cdots, 
	\begin{bmatrix}
	 P_1\\	 S_1\\	 T_1\\	 P_1\\	 S_1\\	 \color{red}{T_1}
	\end{bmatrix}, 
	\begin{bmatrix}
	 \color{red}{P_2}\\	 S_1\\	 T_1\\	 P_1\\	 S_1\\	 T_1
	\end{bmatrix}, 
	\begin{bmatrix}
	 P_2\\	 \color{red}{S_2}\\	 T_1\\	 P_1\\	 S_1\\	 T_1
	\end{bmatrix}, 
	\begin{bmatrix}
	 P_2\\	 S_2\\	 \color{red}{T_2}\\	 P_1\\	 S_1\\	 T_1
	\end{bmatrix}, 
	\cdots, 
	\begin{bmatrix}
	 P_L\\	 S_L\\	 T_L\\	 P_L\\	 \color{red}{S_L}\\	 T_{L-1}
	\end{bmatrix}, 
	\begin{bmatrix}
	 P_L\\	 S_L\\	 T_L\\	 P_L\\	 S_L\\	 \color{red}{T_L}
	\end{bmatrix}. 
\end{align*}
\end{linenomath*}
The transition rule for the lead-lag transformation is as follows:
       \begin{enumerate}
	\item Take \textcolor{blue}{two copies} of $X_0$ and use it as the
	      initial condition.
	\item Update \textcolor{red}{only $1$ component} among $d$ components at once.
	\item Use the previous value instead if the present value is missing.
       \end{enumerate}

  \subsection{Machine learning procedure for quality control process}
Suppose we have a set of profile data
$ X(m)\define \{X_{\tau}(m)\}_{0 \le \tau \le t}$
for $m=1,2,\cdots,M$,  
whose signature is denoted as 
$\mathrm{X}(m)\define \mathcal{S}(X(m))$.
Let us consider the problem of assigning the discriminant values to each profile
depending on whether a profile matches the quality standard.
       \begin{enumerate}
	\item We first make a model for the rule of quality control 
	      as a functional form;
	      that is, 
a linear combination of the 
    iterated integrals $\mathrm{X}^I$ for all combinations of 
    indices $I=(),(i_1), (i_1 i_2), \cdots,  (i_1 \cdots i_n)$
    yields the discriminant value.
    \begin{linenomath*}\postdisplaypenalty=0
      \begin{align}
        y=\sum_{I} w^{I} \mathrm{X}^I+\epsilon,
        \label{linear}
      \end{align}
    \end{linenomath*}
    where $\epsilon$ is the error. 
    Since each index in $I$ runs over $1,2,\cdots,d$,
    the sequence of iterated integrals in Eq.\,(\ref{linear}) has 
    $\sum_{j=0}^6 d^{\bullet j}=(d^{\bullet n+1}-1)/(d-1)$ terms.
    Note that $\mathrm{X}^{()}$ represents the constant $1$.	      
	      Such a representation is possible because its nonlinearity
	      is unraveled thanks to the property of shuffle product;
	      for a fixed path $X$,
	      the product of iterated integrals for indices $A$ and $B$
	      is expressed by the iterated integral with respect to 
	      the shuffle product $A \shuffle B$:
	      \begin{linenomath*}\postdisplaypenalty=0
	       \begin{align}
	       \mathrm{X}^A \mathrm{X}^B &=  \mathrm{X}^{A \shuffle B}. 
	       \end{align}
	      \end{linenomath*}
	      For example,
	      $
	      \mathrm{X}^{(aa)}
	      \mathrm{X}^{(b)}
	      =
	      \mathrm{X}^{(baa)}+  \mathrm{X}^{(aba)}+  \mathrm{X}^{(aab)}
	      $. 
	      This means that a product of iterated integrals is always
	      reduced to the sum of higher-order iterated integrals.
	      Moreover,
	      by virtue of the Stone--Weierstrass theorem,  
	      any nonlinear function of the shape of a path
	      can be represented as a linear combination of the iterated integrals.
	\item 
Suppose we have pairs $(X(m), y(m))$, where 
each $X(m)$ is a profile sequence, and $y(m)=0,1$ is the discrimination value, 
which is already given to each sample $m=1, 2, \cdots, M$ as training data.
Learning these data is simply deriving the weights $w^{I}$ 
that minimize an $L_1$-regularized cost function:
	      \begin{linenomath*}\postdisplaypenalty=0
	       \begin{align}
	       J(w)&=\frac1{2 M}\sum_{m=1}^M \left( y(m) -\sum_{I} w^{I} \mathrm{X}(m)^I \right)^{\bullet 2}
	       +\alpha \sum_{I} \left| w^{I} \right|.
	       \label{cost}	
	       \end{align}
	      \end{linenomath*}
Because the terms in $\sum_I|w^I|$
    are not quadratic but linear,
    they have the effect of selecting significant terms 
    under the summation over the set labeled by $I$.
    This is the notion of least absolute shrinkage and selection operator (LASSO)\shortcite{10.2307/2346178}
    , which can help prevent overfitting.
    The larger the value of $\alpha$ is, the smaller the number of selected terms with $w^I\neq 0$ is.
    To set an appropriate number of terms,
    several values of $\alpha$ will be tested.

    \item
Using the coefficients $w$ derived in (\ref{cost}),
and substituting into Eq\,(\ref{linear}) 
the iterated integrals for a profile
not used for training,
we obtain $\tilde{y}$, an estimate for $y$, as follows.
        \begin{linenomath*}\postdisplaypenalty=0
          \begin{align}
            \tilde{y}(m)&\define \sum_{I} w^{I} \mathrm{X}(m)^I,
          \end{align}
        \end{linenomath*}
        where $w^I$'s are estimated from the minimization of cost\,(\ref{cost}).
	\item The minimization problem is efficiently 
	      solved by the coordinate descent (CD) method 
	      \cite{friedman2007pathwise}.
       \end{enumerate}

       For the $L_1$-regularization term to apply evenly,
each iterated integral $\mathrm{X}^I$ is preprocessed by
subtracting the ensemble mean $\mu^I_{\text{train}}$ of the training ensemble and dividing by the standard 
deviation $\sigma^I_{\text{train}}$ of the training ensemble:
\begin{linenomath*}\postdisplaypenalty=0
 \begin{align}
\mathrm{X}(m)^I &\leftarrow
\frac{\mathrm{X}(m)^I-\mu^I_{\text{train}}}{\sigma^I_{\text{train}}}.
 \end{align}
\end{linenomath*}
The same operation is performed
for the iterated integrals in cross-validation.
The minimization problem is solved by using the Python library {\tt scikit-learn} \cite{scikit-learn}.

\subsection{Assessment of learning results}
The performance of the binary classifier can be 
quantitatively assessed by 
visualizing it with the receiver operating characteristic (ROC) curve \cite{egan1975signal}.
We refer to the profiles that pass the quality criterion 
as negative (normal) $y=1$, and the others as positive (bad) $y=0$.
By shifting the cutoff value $y_c$,
one can count the number of positive ones with $\tilde{y}< y_c$
and that of negative ones with $y_c \leq \tilde{y}$.
Then, the samples fall into the four categories in Table \ref{cm}.
\begin{table}
\begin{center}
  \caption{Confusion matrix with cutoff $y_c$
  \label{cm}}
  \begin{tabular}{l||r|r}
    \backslashbox{Estimated $\tilde{y}$}{True $y$}      & $0$ & $1$\\
    \hline
    $\tilde{y} < y_c$ & True-positive $N_{\text{TP}}$ & False-positive $N_{\text{FP}}$\\
    $y_c \le \tilde{y}$ & False-negative $N_{\text{FN}}$ & True-negative$ N_{\text{TN}}$
  \end{tabular}
\end{center}
\end{table}

The true-positive rate is defined as
	      $N_{\text{TP}}/(N_{\text{TP}}+N_{\text{FN}})$,  
and the false-positive rate as
	     $N_{\text{FP}}/(N_{\text{FP}}+N_{\text{TN}})$.
The ROC curve is the two-dimensional
plot of false-positive rate versus true-positive rate,
by changing the cutoff $y_c$.
It has better performance if the trajectory approaches the 
upper left corner.
Therefore, the area under the ROC curve indicates 
the performance.

Note that, to improve the readability of the histograms,
        we use a modified estimation value:
        \begin{linenomath*}\postdisplaypenalty=0
          \begin{align}
            \tilde{y}(m)&\define 1-\left|1-
            \sum_{I} w^{I} \mathrm{X}(m)^I\right|,
          \end{align}
        \end{linenomath*}
        where we apply transformation $y \mapsto 1-|1-y|$
        so that $\tilde{y}(m)\leq 1$.
\subsection{Experiment using PCA}
Alternatively, 
  principal component analysis (PCA) \cite<e.g.,>{thomson2014}
  can be applied to represent the normal profiles.
  In that case, the experiment for estimating 
  the quality control flag is performed as follows.
  \begin{enumerate}
  \item 
    We apply the same nondimensionalization 
    to the $T$- and $S$-sequences as the signature method,
    and then perform nearest-neighbor interpolation 
    at points $2000 \widehat{P} =5, 15,\cdots, 1995,$
    which are placed every $10\unit{dbar}$.
    Accordingly, the sequences are transformed into 
    a sequence $X=(X_1,\cdots,X_L)^T$ with $L=400$.
  \item
    Let 
    $\mathcal{F}\define\left\{X(m)|m=1,2,\cdots,M\right\}$ be the set of all profiles.
    We randomly choose
    the training ensemble $\mathcal{F}_{\text{tr}}\subset \mathcal{F}$,
    which comprises negative (normal) samples 
    $\mathcal{F}_{\text{tr},n}\subset \mathcal{F}_{\text{tr}}$,
    and positive (bad) samples 
    $\mathcal{F}_{\text{tr},p}\subset \mathcal{F}_{\text{tr}}$.
  \item 
    Training is performed  by computing the principal components (PCs)
    for negative training samples $\mathcal{F}_{\text{tr},n}$.
    Let $$
    U=\begin{bmatrix}
    U_1^1&\cdots&U_1^L\\
    \vdots&&\vdots\\
    U_{N_{\text{pc}}}^1&\cdots&U_{N_{\text{pc}}}^L
    \end{bmatrix}$$
    be the truncated PCs, and 
    $\overline{X}=(\overline{X}_1,\cdots,\overline{X}_L)^T$ 
    be the ensemble mean for the training ensemble
    $\mathcal{F}_{\text{tr},n}$.
  \item
    For the $m$-th profile $X(m)\in \mathcal{F}_{\text{tr}}$ 
    (or $\in \mathcal{F} \setminus \mathcal{F}_{\text{tr}}$ for cross-validation),
    the mean square residual for
    representing it by the first $N_{\text{pc}}$-PCs is computed as
    \begin{linenomath*}\postdisplaypenalty=0
      \begin{align}
        r(m)&= \frac{1}{2L}
        \left|(I_L-U^T U)(X(m)-\overline{X})\right|^{\bullet 2},
      \end{align}
    \end{linenomath*}
    where $I_L$ is an $L$-dimensional identity matrix.
    The estimated $\tilde{y}$ is thereby defined as
    $\tilde{y}(m)\define -r(m).$
  \item
    For a fixed threshold value $y_c$, 
    we assign negative to the $m$-th profile if $\tilde{y}(m)>y_c$
    and positive otherwise.
    The ROC curve is drawn by plotting false-positive rates versus true-positive 
    rates for  various $y_c$.''
  \end{enumerate}

\section{Results and Discussion}
\begin{figure}
  \begin{center}
   \includegraphics[width=0.95\columnwidth]{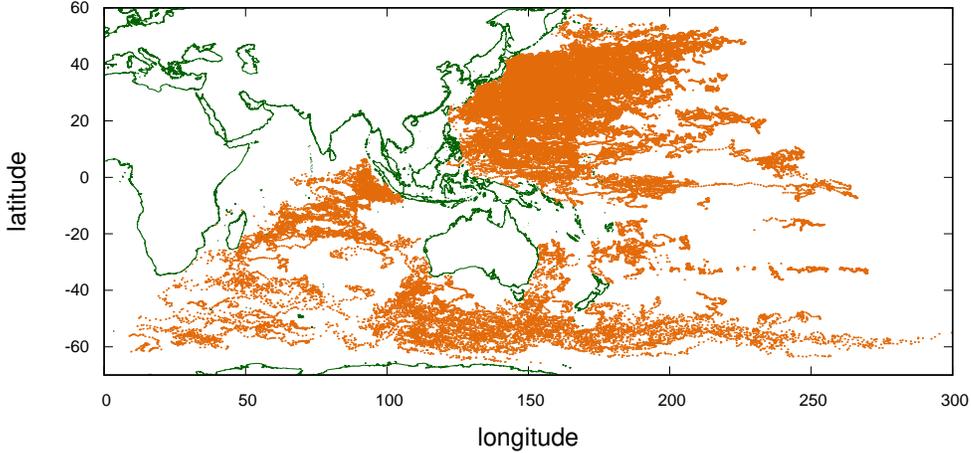}
  \caption{Location of the Argo profiles used 
   in this study.\label{map}}
  \end{center}
\end{figure}

We used a dataset 
    observed at the location shown in Fig.\,\ref{map}.
    Each profile is assigned a delayed-mode QC flag 
    by Japan Agency for Marine-Earth 
    Science and Technology (JAMSTEC).
    We treated profiles with depth widths (the difference between the minimum and
    maximum depths) of more than $1000 \mathrm{m}$,
    and each profile had approximately $L\sim 100$ observation points.
    The number of profiles was $M=8.2\times 10^4$, and
    the training data were randomly chosen from these profiles.
    After applying the lead-lag transformation,
    each profile was converted into the signature up to order $n=6$.

An overview of the machine learning 
results is shown by the 
histogram of estimated values $\tilde{y}$ for 
normal samples ($y=1$), and the histogram 
for bad samples ($y=0$).

\begin{figure}
  \begin{center}
   \includegraphics[width=0.8\columnwidth]{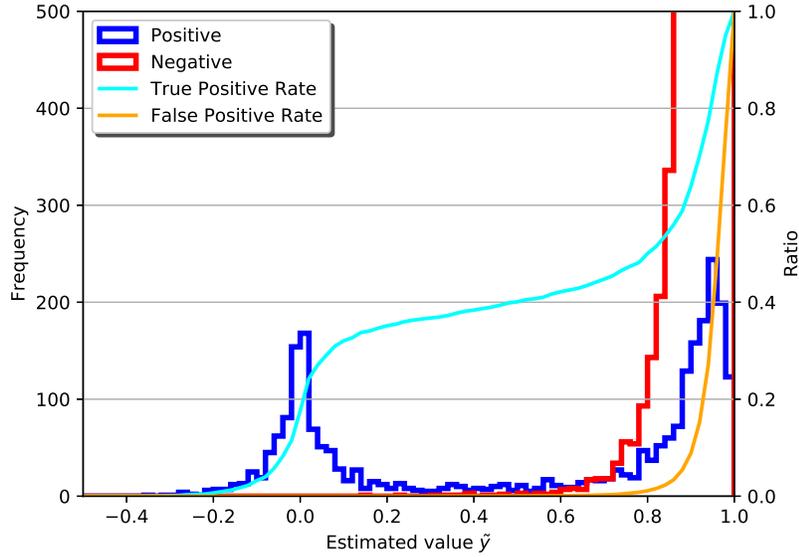}
  \caption{
Histogram for discriminant analysis (identification of the training data).
 Proportion of training data: $0.4$,  $\alpha=10^{-5}$.
The horizontal axis is the estimated value $\hat{y}$,
      and the vertical axis is its frequency.
      Blue: histogram for positive data (bad samples with flag $y=0$),
      red: histogram for negative data  (normal samples with flag $y=1$),
     cyan: true-positive rate, and
     orange: false-positive rate.
    \label{hist_0.4t}}
  \end{center}
\end{figure}

\begin{figure}
  \begin{center}
   \includegraphics[width=0.8\columnwidth]{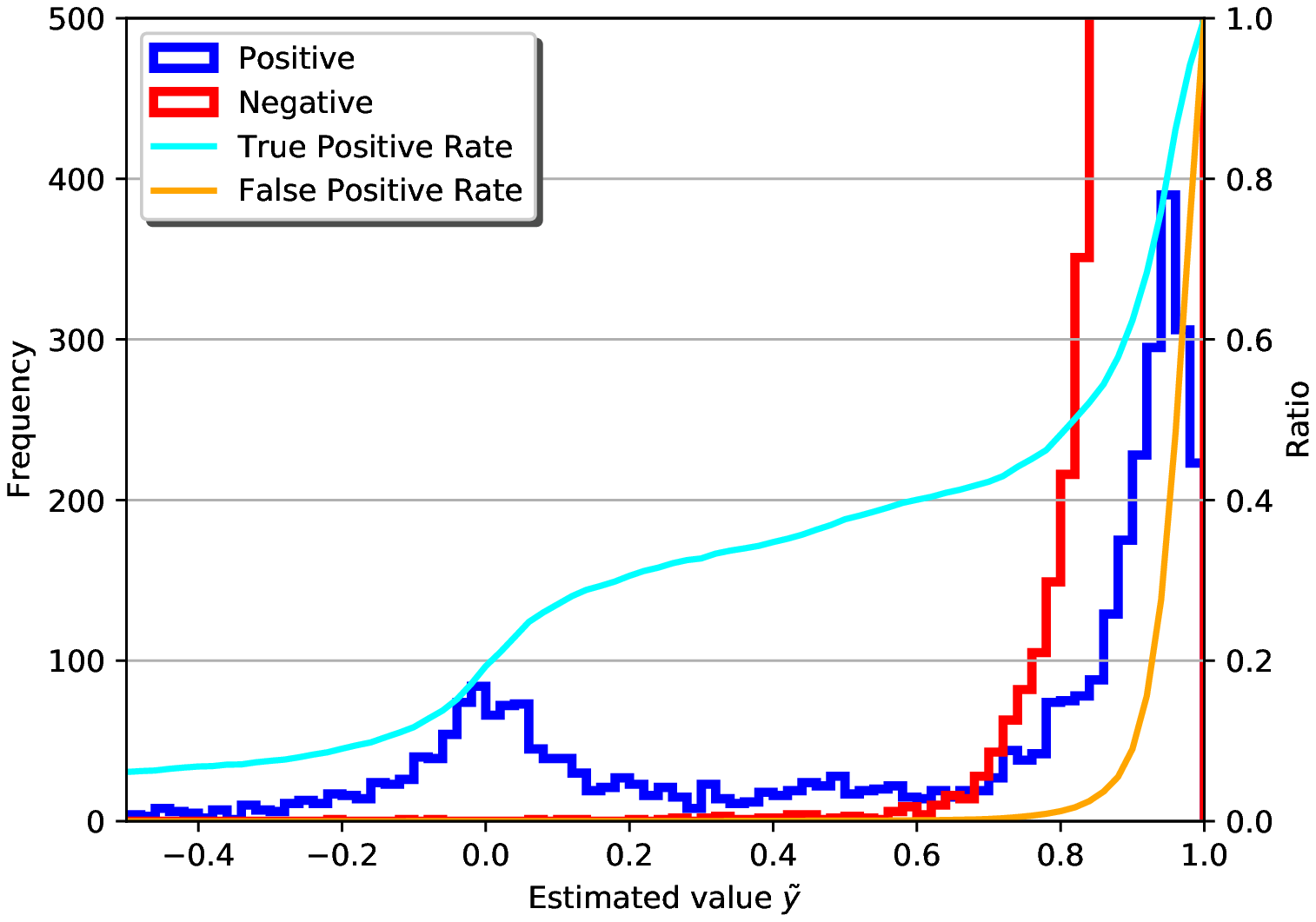}
  \caption{
Histogram for discriminant analysis (cross-validation).
Proportion of training data: $0.4$,  $\alpha=10^{-5}$.
The horizontal axis is the estimated value $\hat{y}$,
      and the vertical axis is its frequency.
      Blue: histogram for positive data (bad samples with flag $y=0$),
      red: histogram for negative data  (normal samples with flag $y=1$),
     cyan: true-positive rate, and 
     orange: false-positive rate.
    \label{hist_0.4c}}
  \end{center}
\end{figure}

\begin{figure}
  \begin{center}
   \includegraphics[width=1.0\columnwidth]{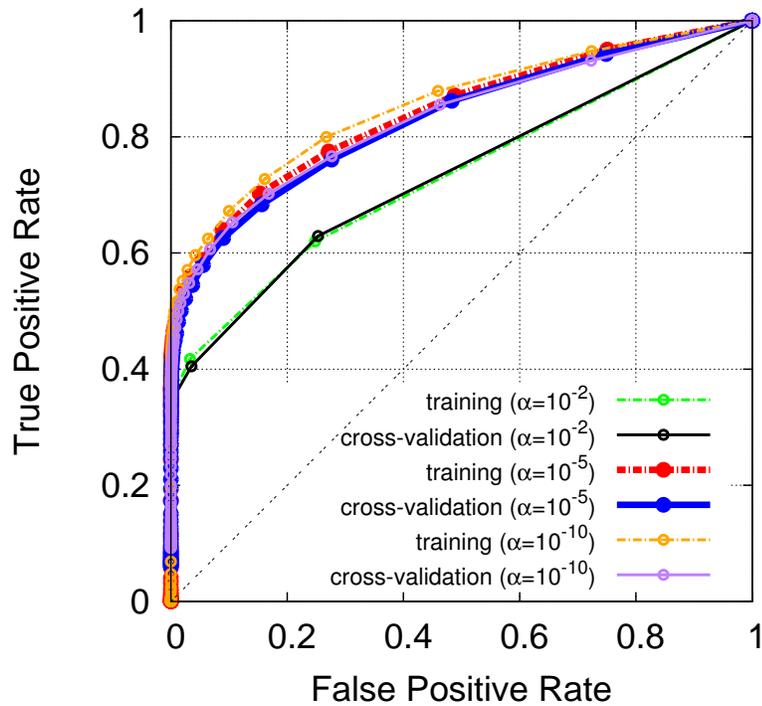}
  \caption{ROC curves for various regularization parameters.
   Identification of the training data (broken curves)
   and cross-validation (solid curves) are depicted.
            The horizontal axis is the false-positive rate, and 
	the vertical axis is the true-positive rate.
    \label{roc}}
  \end{center}
\end{figure}

Figures \ref{hist_0.4t} and  \ref{hist_0.4c} show the histograms
when $40\%$ of the data are used for training
and the remaining $60\%$ are used for cross-validation.
We can see that learning is properly performed because
there is little difference between the 
identification of training data and the cross-validation.
In particular, this approach never misidentifies negative (normal) profiles
if the appropriate cutoff $y_c$ is used,
but it may accept positive (bad) profiles with a probability $0.6$ when $y_c=0.5$.
This property is also reflected in the tendency of   
the ROC curve (Fig.\,\ref{roc}) to be almost tangent to the
horizontal axis when $x$ is small, but not tangent to $y=1$ when $y$ is large.
The histogram for positive samples has two clear peaks, 
which suggests that the ambiguity is not caused by 
the judgment by the machine learning,
but by the fact that the original quality control flag had
a criterion that cannot be decided only by the shape.
For example, the original quality control, encoded in $y$, may have
      a criterion about deviation from climatological variation, which 
      cannot always be detected from profile shape.
      Moreover, the original quality control is partly done through visual checking, 
      for which the criteria can fluctuate between checks. 
      Obviously, both are not represented by the signature, which 
      is static and shape-oriented.

\begin{figure}
  \begin{center}
   \includegraphics[width=0.8\columnwidth]{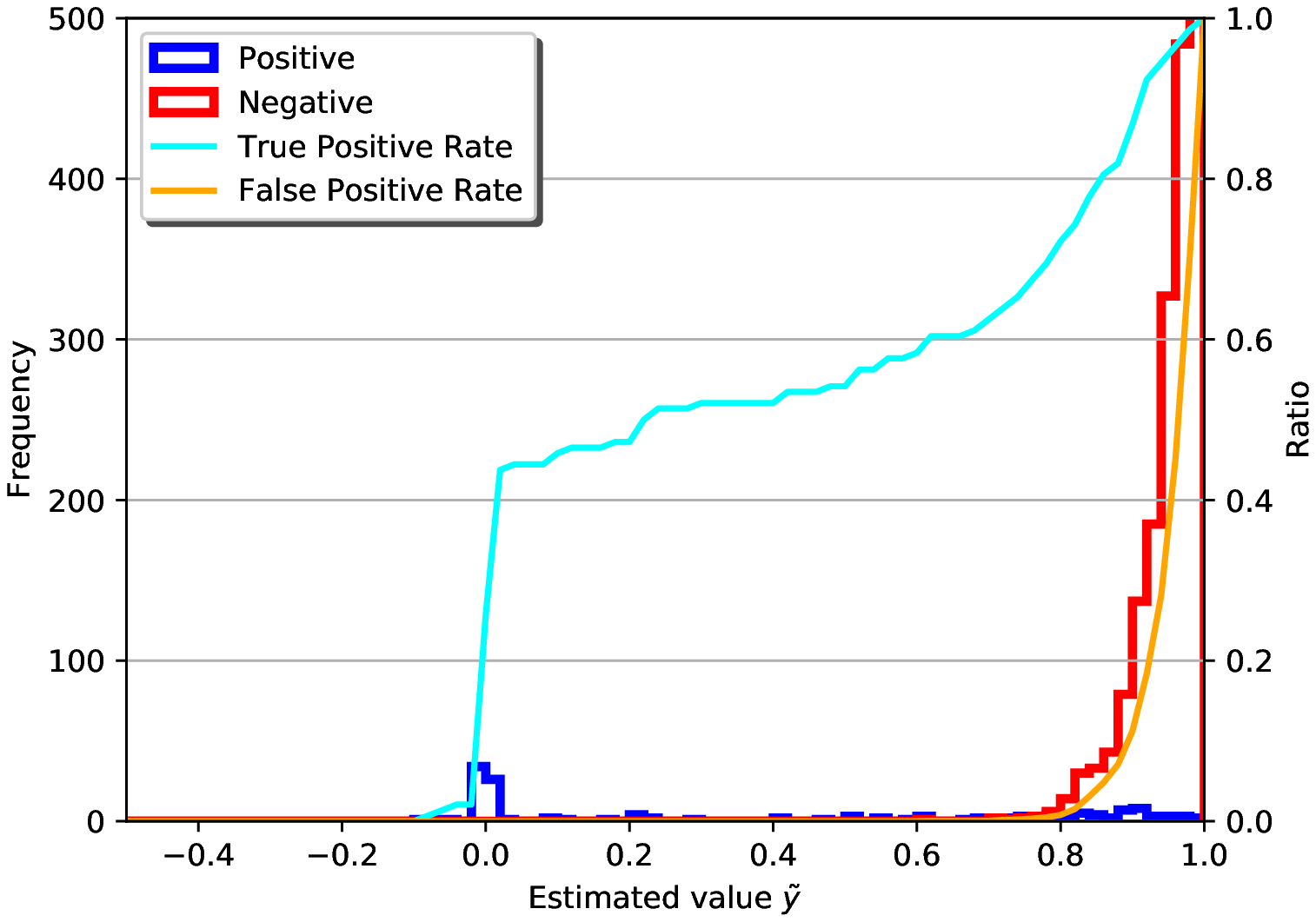}
  \caption{
Histogram for discriminant analysis (identification of the training data).
Proportion of training data: $0.025$,  $\alpha=10^{-5}$.
The horizontal axis is the estimated value $\tilde{y}$,
and the vertical axis is its frequency.
   Blue: histogram for positive data (bad samples with flag $y=0$),
     red: histogram for negative data  (normal samples with flag $y=1$),
     cyan: true-positive rate, and
     orange: false-positive rate.
    \label{hist_0.025t}}
  \end{center}
\end{figure}

\begin{figure}
  \begin{center}
   \includegraphics[width=0.8\columnwidth]{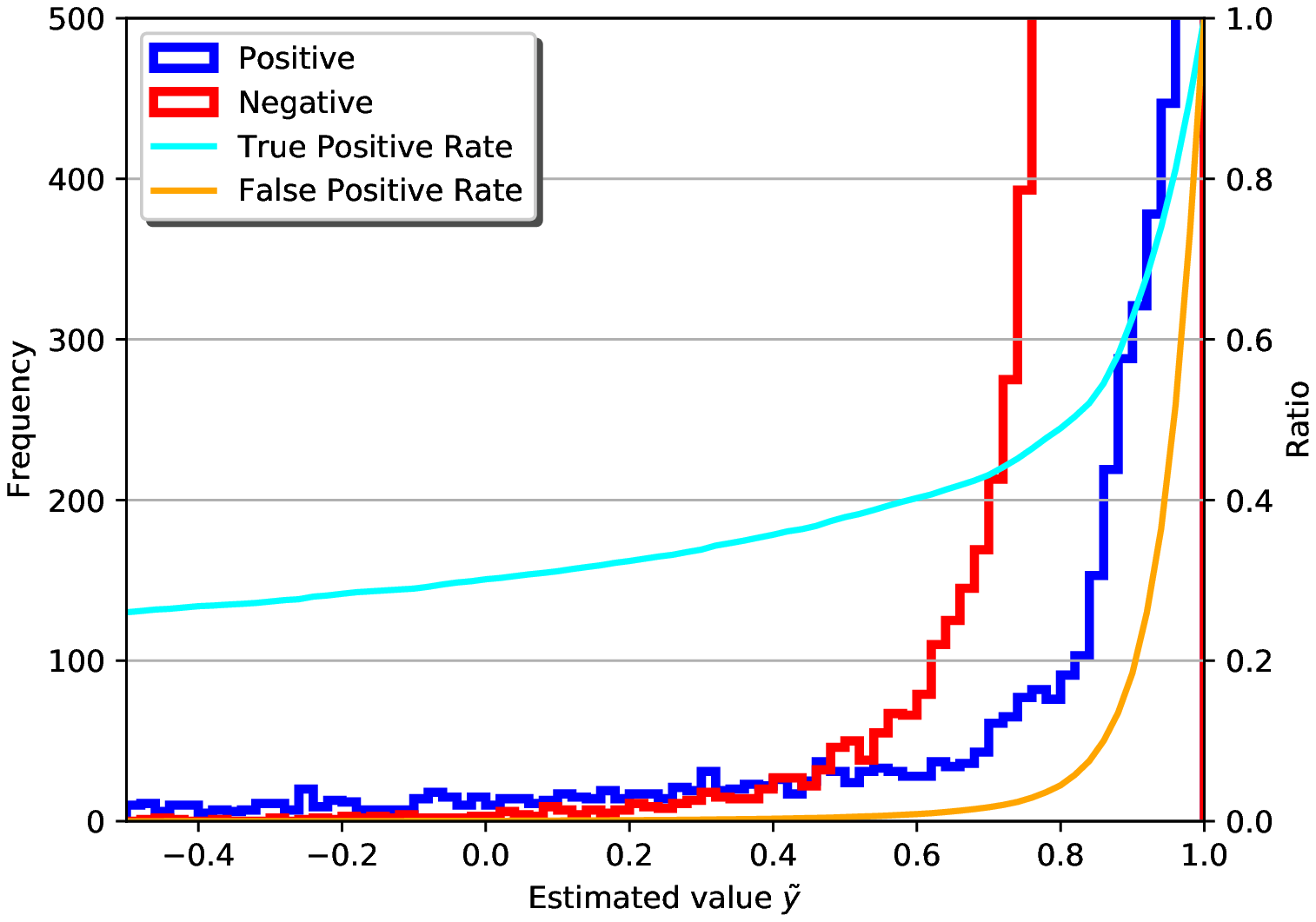}
  \caption{
Histogram for discriminant analysis (cross-validation).
Proportion of training data: $0.025$,  $\alpha=10^{-5}$.
The horizontal axis is the estimated value $\tilde{y}$,
and the vertical axis is its frequency.
   Blue: histogram for positive data (bad samples with flag $y=0$),
     red: histogram for negative data  (normal samples with flag $y=1$),
     cyan: true-positive rate, and 
     orange: false-positive rate.
    \label{hist_0.025c}}
  \end{center}
\end{figure}

Figures \ref{hist_0.025t}  and \ref{hist_0.025c} shows the histograms
when $2.5\%$ of the data are used for training
and the remainder is used for cross-validation.
In this case, there is a clear tendency of over-learning, 
which indicates that the number of learning samples, $2.5\%$, is not
sufficient.

\begin{figure}
  \begin{center}
   \includegraphics[width=1.0\columnwidth]{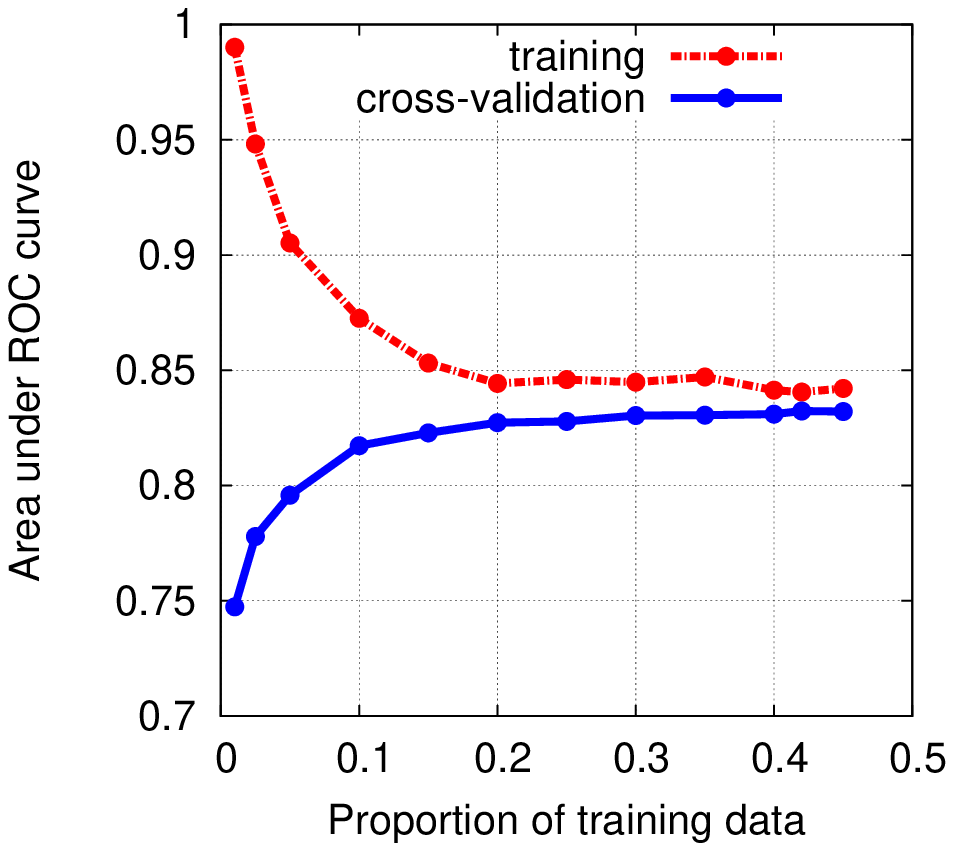}
 \caption{
Learning curves for different proportions of training data.
 Red: identification of the training data; 
 blue: cross-validation.
    \label{l-curve}}
\end{center}
\end{figure}

The performance of a method can be measured 
      from the area under the ROC curves (AUC).
Comparing that for the experiments with various ratios of
learning samples, we notice that 
over-learning occurs when the ratio is less than $20\%$
(Fig.\,\ref{l-curve}).  

\begin{figure}
  \begin{center}
   \includegraphics[width=1.0\columnwidth]{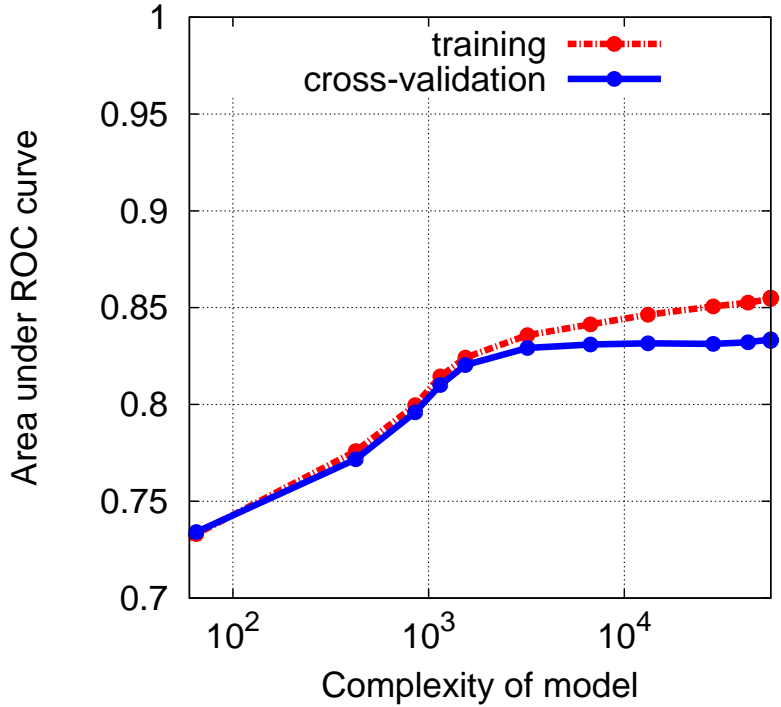}
 \caption{
Learning curves for different model complexities.
 Red: identification of the training data; 
 blue: cross-validation.
    \label{l-curve_alpha}}
  \end{center}
\end{figure}

We also
compared the results of the experiments
with various weights $\alpha$ of the regularization term
by the AUC.
The number of terms 
under the summation over the set labeled by $I$ in Eq.\,(\ref{cost})
is dependent on $\alpha$.
Therefore, if we increase the degrees of freedom of the coefficients $w$ by 
using a smaller $\alpha$, the performance of the reproduction capability
increases.
However, the estimation capability begins to
saturate at approximately $6700$ 
degrees of freedom (Fig.\,\ref{l-curve_alpha}),
where $\alpha=10^{-5}$ is used.
At that point, the complexity seems to become appropriate.

\begin{figure}
  \begin{center}
   \includegraphics[width=1.0\columnwidth]{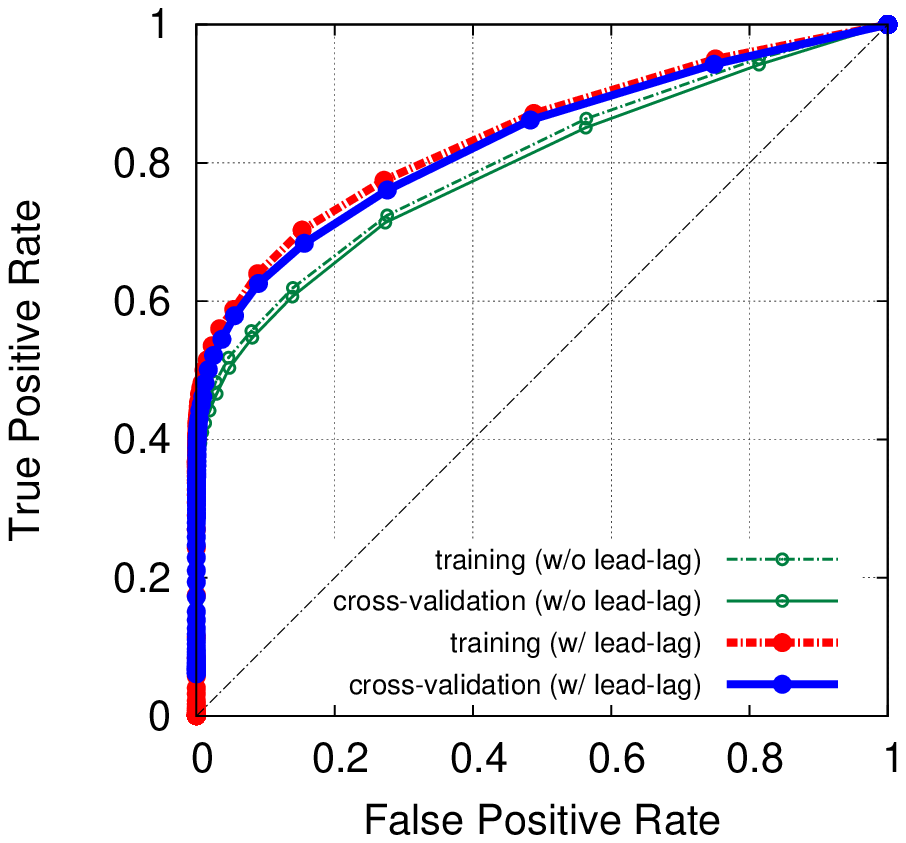}
  \caption{ ROC curves for the cases with and without lead-lag transformation.
        The case with lead-lag transformation (red and blue) is compared to 
        that without lead-lag transformation (green);
        identification of the training data (broken curves)
        and cross-validation (solid curves) are depicted.
        The horizontal axis is the false-positive rate, and 
	the vertical axis is the true-positive rate.
    \label{roc_woll}}
  \end{center}
\end{figure}

To confirm the efficacy of lead-lag transformation,
    we performed a similar experiment as in Figs.\,4 and 5
    except without lead-lag transformation.
    We set $\alpha=10^{-5}$ and the proportion of training data to $0.4$.
    Figure\,\ref{roc_woll} shows the ROC curves for the experiment.
    The curves for the case with lead-lag are on the upper-left of 
    those for the case without lead-lag,
    which indicates that the lead-lag transformation helps improve
    the estimation of the quality control flag.

\begin{figure}
  \begin{center}
   \includegraphics[width=1.0\columnwidth]{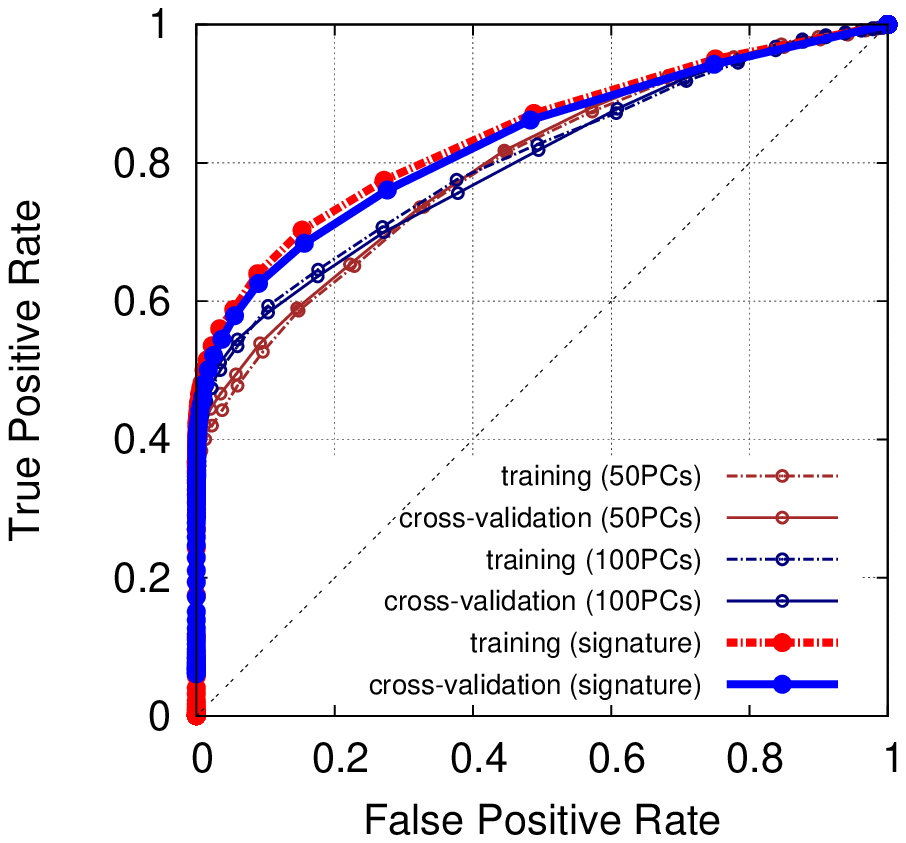}
  \caption{ ROC curves for the cases using signature and PCA.
      The case using the signature method (red and blue) is compared to the 
      ones using the PCA method with $50$PCs (brown) and $100$PCs (purple);
      identification of the training data (broken curves)
      and cross-validation (solid curves) are depicted.
      The horizontal axis is the false-positive rate, and 
      The vertical axis is the true-positive rate.    \label{roc_eof}}
  \end{center}
\end{figure}

As a reference case using another representation of the shape,
  we performed PCA experiments with $N_{\text{pc}}=50$ and $100$PCs.
  The proportion of training data is set to $0.4$, 
  the same as for the signature case.
  Figure\,\ref{roc_eof} depicts the ROC curves for the PCA experiments.
  Although the PCA method also exhibits a considerable skill,
  the curves stay to the lower-right of those for the signature method,
  which indicates that the signature method is more effective than the PCA method
  in estimating the quality control flag.
  Meanwhile, the computational cost for the signature method is 
  not significantly higher than that for the PCA method,
  because the former only additionally requires converting each data sequence into the truncated signature, 
  whose calculation load is low.

\begin{figure}
  \begin{center}
   \includegraphics[width=1.0\columnwidth]{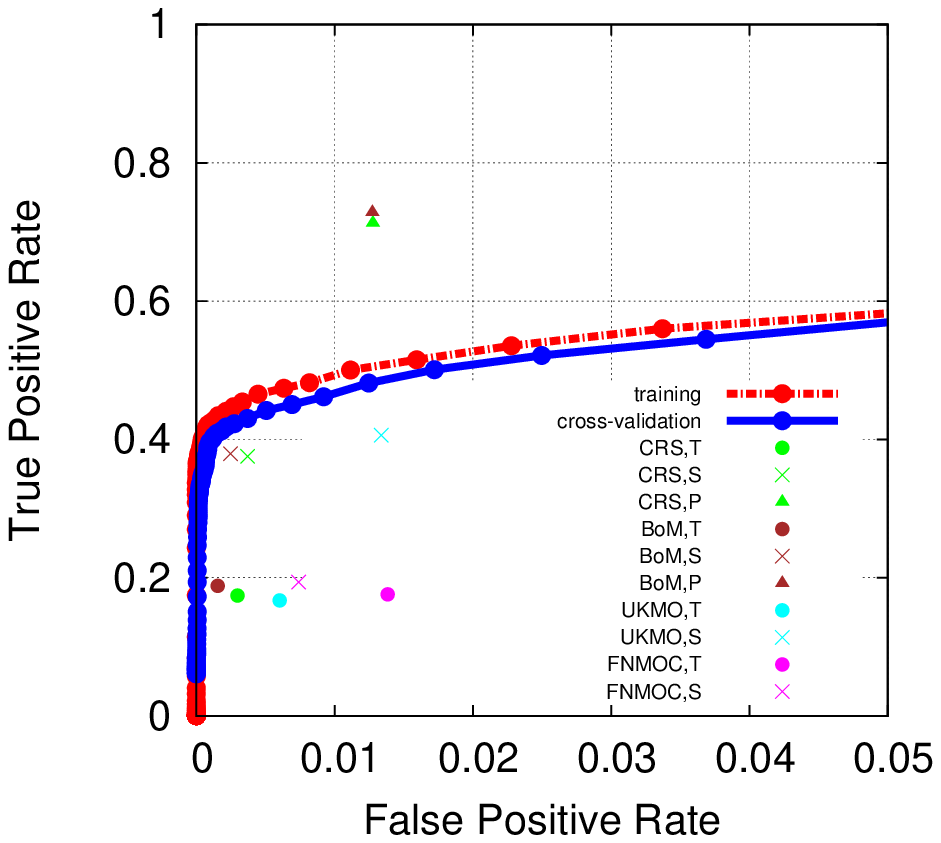}
  \caption{ Performance points for real-time QC on ROC graph. 
      The red and blue curves are for the signature method.
      The points denote
      temperature (circle), salinity (cross), and pressure (triangle), 
      each of which represents the false-positive vs true-positive rates
      for real-time QC data from 
      the Coriolis data center (CRS; green),
      the Australian Bureau of Meteorology (BoM; brown),
      the United Kingdom Met Office (UKMO; cyan), and
      the Fleet Numerical Meteorology and Oceanography Centre (FNMOC; magenta),
      taken from Table\,5 of \protect\shortciteA{wedd2015argo}.
    \label{roc_rqc}}
  \end{center}
\end{figure}

Further, comparison with the data from the ARGO intercomparison project is performed.
  The performances of the real-time assignment of QC flags \shortcite{wong2020argo} 
  by several institutes are shown in \shortciteA{wedd2015argo},
  when the corresponding assignments by the delayed-mode QC are regarded
  as the ground truth.
  Because the sets of profile data differ from those in our case,
  a direct comparison is not strictly relevant 
  but will still serve as a measure of the performance.
  Figure\,\ref{roc_rqc} shows the false-positive vs true-positive rates
  for those samples in comparison to the signature case.
  Apart from the the pressure data,
  all the points for real-time QC data lie on the bottom-right side of 
  our ROC curve. 
  This suggests that the signature method may assign 
  the QC flags more efficiently than the real-time QC procedure does,
  provided that the past assignment results are ready for use.
  Another advantage of the signature method is that it 
  assigns the flags consistently to all the components $(P,S,T)$ with
  higher reliability.

Overall, we found that
machine learning using the signature method
can learn the existing quality control flags of Argo profiles and
automatically assign the flag to new profiles,
but 
it sometimes overlooks bad samples
because of the ambiguity inherent in the original quality control flag.
Comparative study shows the signature method has a higher performance for estimating 
the flag than other conventional methods, 
including the one with the PCA representation 
and the operational assignments of real-time QC.


\section{Conclusions}
In this research, 
we first demonstrated that the shape of a profile from the Argo ocean observing array 
 can be represented by the iterated integrals.
Then, we constructed a model for the function that assigns
a quality control flag to the shape of a profile,
which is expressed as a weighted sum of the iterated integrals.

We performed supervised learning for the weights 
using the existing quality control flags for training data,
and demonstrated via 
cross-validation that it has good performance in estimating flags for unknown data.

A comparative experiment using the PCA method showed
    that the signature method, in combination with lead-lag transformation, 
    outperforms the PCA method
    in estimating the quality control flag.
    This suggests the superiority of the signature method
    compared to the conventional machine-learning technique.

This algorithm can potentially
enable automatic assignment of quality control 
flags to new Argo data.
The significance of the algorithm is that it 
objectively and automatically assigns the quality control flag
only on the basis of past knowledge about the
quality of data without imposing any ad hoc rules.
Hence, 
it should enable 
more objective and efficient quality control
compared to traditional manual methods or rule-based machine learning.

The signature method is quite effective 
for expressing the 
shape of an Argo profile and its nonlinear function quantitatively.
The rationale for this advantage is that a 
nonlinear and complicated function of assigning quality control flags
can be transformed into a linear combination of the iterated integrals
through algebraic transformation (shuffle product)
without introducing any errors.
This is superior to conventional 
multivariate regression models, which approximately 
regard nonlinear dependencies as linear ones.
Along this line, we can express,
as a function of signature,
not only quality control flags but also any oceanic phenomena.

One application of the signature method
is assimilation of the signature of observational data 
into a general ocean circulation model.
For example,
we can convert a vertical sequence of observational data and 
that of model data into iterated integrals.
We then construct a cost function that compares the signatures for model
and observation,
rather than directly comparing the state vectors 
composed of temperature and salinity at each depth.
Although a cost term is for a single horizontal and temporal point,
data assimilation, in particular the four-dimensional variational method,
can combine the effects from multiple terms via model
integration and adjoint integration.
By doing so, we gain the advantage that 
the projection of a vertical profile onto any ocean phenomena 
attains a linear form, which 
will result in efficient data assimilation.
This is expected because
many diagnoses for oceanic conditions 
are written in terms of iterated integrals,
as illustrated in sec.\,\ref{TB} and \ref{da}.

\appendix 
\section{Picard iteration\label{Pic}}
To understand the notion of signature, consider
how the theory of rough path treats a data sequence acting on a system.
Suppose we have a system of ordinary differential equations with respect
 to $Y_{\tau}$ forced by a path $X_{\tau}$:
\begin{linenomath*}\postdisplaypenalty=0
\begin{align}
  dY^i_{\tau} &= \sum_{j,k} F^i_{jk} Y^j_{\tau} dX^k_{\tau},\label{differential}
\end{align}
\end{linenomath*}
where 
$Y^j_{\tau}$ is
the $j$-the component of vector $Y_{\tau}$, 
and $F^i_{jk}$ is the $i,j,k$-th component of $3$-dimensional tensor $F$.

Performing the Picard iteration yields a solution:
\begin{linenomath*}\postdisplaypenalty=0
\begin{align}
Y^i_t
&=
\sum_{n=0}^{\infty}
\sum_{i_{\cdot},j,k_{\cdot}}F^i_{i_{n-1}k_n}
\cdots
F^{i_2}_{i_1k_2}
F^{i_1}_{jk_1} \mathrm{X}_n^{(k_1k_2\cdots k_n)}Y_0^j,
\end{align}
\end{linenomath*}
where
$\mathrm{X}_n^{(k_1k_2\cdots k_n)} \define
\int_{0<\tau_1< \cdots < \tau_n<t}
dX_{\tau_1}^{k_1} dX_{\tau_2}^{k_2} \cdots  dX_{\tau_n}^{k_n}$
is a component of
the $n$-th iterated integral (\ref{IteratedI}).
By omitting the indices,
we can simply write the solution as $Y_t=\left[\sum_{n=0}^{\infty}F^{\otimes n} \mathbf{X}_n\right] Y_0.$
Notice that the convergence of the series is guaranteed because 
the magnitude of each iterated integral
is uniformly bounded: $|\mathrm{X}_n^{(k_1k_2\cdots k_n)}|<\frac{L^{\bullet n}}{n!}$, where
$L$ is the path length.
This form of solution suggests that
the action of $X$ on $Y$ can be well summarized by the iterated integrals, and
an approximate solution is reproduced by 
a truncated series of iterated integrals $(\mathbf{X}_0,\mathbf{X}_1,\cdots,\mathbf{X}_n)$,
which is called a truncated signature up to order $n$.
The point is that the effect of a forcing on a system 
is asymptotically approximated by
the truncated path signature
but not by the partial sequence of state vectors.


It has been proven that a
path that never crosses itself, 
like in the case of Argo profiles,
is completely determined by its signature \cite{hambly2010uniqueness}.
A function of a path, say $\phi$, 
can thus be regarded as
that of its signature and compactly approximated 
by that of a truncated signature:
\begin{linenomath*}\postdisplaypenalty=0
\begin{align}
\phi\left(\{X_{\tau}\}_{0\leq \tau \leq t}\right)
&\fallingdotseq f(\mathbf{X}_0,\mathbf{X}_1,\cdots,\mathbf{X}_n).
\end{align}
\end{linenomath*}
A further advantage of such treatment is that 
the function $f$ can always be  
expressed as a linear combination of 
iterated integrals,  
owing to the shuffle-product
property, which is explained later.
\section{Thermal wind flow in terms of iterated integrals\label{da}}
As an example of higher-order iterated integrals,
we show here that thermal wind flow can be written with iterated integrals.

The thermal wind relation is written in vertical $P$-coordinates as
  \begin{equation}
f\frac{\partial u}{\partial P} = -\left.\frac{\partial}{\partial y}(\rho^{\bullet -1})\right|_{P=\text{const.}},\quad
f\frac{\partial v}{\partial P} = \left.\frac{\partial}{\partial x}(\rho^{\bullet -1})\right|_{P=\text{const.}},
  \end{equation}
where $f$ is the Coriolis parameter, 
$u,v$ are velocity,
$\rho$ is density, and $x,y$ are the longitudinal and latitudinal coordinates, respectively.
For a fixed latitude $y$, 
by performing integrations along the $x$ direction and then the $P$ direction,
we obtain an estimate for the meridional velocity as
  \begin{linenomath*}\postdisplaypenalty=0
\begin{align}
f\int_{x'=x_0}^{x_1}\frac{\partial v}{\partial P} dx'
&=
\rho(x_1,P')^{\bullet -1}-\rho(x_0,P')^{\bullet -1}
=:\left[\rho(x,P')^{\bullet -1}\right]_{x=x_0}^{x_1},\\
f\int_{x'=x_0}^{x_1}v(x',P) dx'
&=
\left[\int_{P'=P_0}^{P}\rho(x,P')^{\bullet -1}dP'
\right]_{x=x_0}^{x_1},
    \end{align}
\end{linenomath*}
where we set $v(x',P_0)=0$ as the layer of no motion. 
Integrating again along the $P$ direction, we obtain the meridional flow rate as
  \begin{linenomath*}\postdisplaypenalty=0
\begin{align}
Q_v&:= -g^{\bullet -1}
\int_{P''=P_0}^{P_1}
\int_{x'=x_0}^{x_1}v(x',P'') dx'dP''\nonumber\\
&= -(gf)^{\bullet -1}
\left[
\int_{P''=P_0}^{P_1}
\int_{P'=P_0}^{P''}\rho(x,P')^{\bullet -1}dP'
dP''
\right]_{x=x_0}^{x_1},
\label{qv}
    \end{align}
\end{linenomath*}
where the unit is in $\unit{[kg s^{\bullet -1}]}$ because of the $p$-coordinate.

Let $\tau \in [0,1]$ be a 
parameter for the order of observational points in a profile.
Evaluating the density in Eq.\,(\ref{qv})
with the state equation $\varrho$, we have 
  \begin{linenomath*}\postdisplaypenalty=0
\begin{align}
\rho(x,P)^{\bullet -1}&=\varrho\left(T(x,\tau),S(x,\tau),P(x,\tau)\right)^{\bullet -1}
=\varrho\left(\int_0^{\tau} dT_{\tau'},\int_0^{\tau} dS_{\tau'},
\int_0^{\tau} dP_{\tau'}\right)^{\bullet -1},
    \end{align}
\end{linenomath*}
which has iterated integrals as independent variables.
Notice that the shuffle-product property 
transcribes this as a linear combination of iterated integrals.
Substituting this into Eq.\,(\ref{qv}) finally yields
\begin{multline}
 Q_v = -(gf)^{\bullet -1}
\left[
\int_{\tau_3=0}^1
\int_{\tau_2=0}^{\tau_3}
\varrho\left(\int_{\tau_1=0}^{\tau_2}dT_{\tau_1},
\int_{\tau_1=0}^{\tau_2}dS_{\tau_1},\int_{\tau_1=0}^{\tau_2}dP_{\tau_1}
\right)^{\bullet -1}
dP_{\tau_2}
dP_{\tau_3}
\right]_{x=x_0}^{x_1}.
\end{multline}
This shows that the meridional flow rate $Q_v$ is represented as 
a linear combination of iterated integrals 
with respect to $T,S,$ and $P$.

\acknowledgments
The authors appreciate the members of JAMSTEC Argo
data management team for preparing and compiling the Argo profile data.
 All numerical computations were
 performed on the JAMSTEC DA supercomputer system.
Argo float data and metadata are freely available 
from Global Data Assembly Centre 
(Coriolis GDAC \url{http://www.coriolis.eu.org/Observing-the-Ocean/ARGO} or
USGODAE GDAC \url{https://nrlgodae1.nrlmry.navy.mil/argo/argo.html}).
The processing codes are available on Zenodo \cite{git2020}.


%
%

\bibliography{ref}

%
%
%
%
%

\end{document}